# Geophysical signature of the transition zone between the sedimentary cover and the basement: an analogue approach to help de-risking geothermal prospects

M. DARNET[1], C. DEZAYES[1], J. F. GIRARD[2], J.M. BALTASSAT[1], C. LEROUGE[1],

T. REUSCHLE[2], N. COPPO[1], F. BRETAUDEAU[1], J. PORTE[1], Y. LUCAS[3]

[1]BRGM, [2]EOST-IPGS, [3]EOST-LHyGeS

m.darnet@brgm.fr

## ABSTRACT

Within the frame of the ANR-funded CANTARE-Alsace project, we have undertaken a multi-scale and multi-disciplinary approach to increase our knowledge of the transition zone between the sedimentary cover and the basement and provide fundamental knowledge for the assessment of its geothermal potential. In this paper, we report out the results of a study performed on an exhumed transition zone in the Ringelbach area in the Vosges Mountains, on the flank of the Rhine graben. In this analogue of a deeply buried transition zone of the Rhine Graben, a thin layer of Triassic sandstones is still present on the top of the fractured and altered granitic basement providing the opportunity to study in-situ the physical properties of this transition zone. In this paper, we focused on electrical and acoustic properties of the transition zone as they are the main physical parameters usually assessed with the help of geophysical methods during the exploration phase of a geothermal project.

## 1. INTRODUCTION

Exploiting geothermal resources at temperatures between 120 and 200°C in sedimentary and basement rocks, in rifts or in flexural basins, to produce electricity is now possible because of the development of Enhanced Geothermal System (EGS) technology. Reaching such temperature range is a major challenge for mainland France and Europe as it usually requires drilling up to 4 to 6 km depth. Aside from temperature, two other conditions are required to allow the exploitation of the geothermal energy: the presence of fluid, which is the heat vector, and sufficient permeability to allow the production and re-injection of the fluid. This translates into the reservoir being located in the deep layers of sedimentary basins and the upper part of the Paleozoic basement, including the transition zone between the two.

T The reality is however even more complex as the geothermal potential of this zone is also strongly influenced by large heterogeneities including 1) variety and distribution of the lithologies at the cover/basement interface, including products of basement dismantling, and amounts of clay minerals, 2) fracture and fault networks at the transition zone and in the granitic fractured reservoir, 3) other key parameters controlling petrophysical properties at the transition: internal fabrics, paleoweathering and hydrothermal alterations of the granite, and cementation/dissolution in sandstones. It is therefore clear that the characterization of the transition zone and its heterogeneity in the deepest part of sedimentary basins constitutes one of the most challenging problems for the development of geothermal resources.

Within the frame of the CANTARE-Alsace project, we have undertaken a multi-scale and multi-disciplinary approach to increase our knowledge of this transition zone between the sedimentary cover and the basement and provide fundamental knowledge for the assessment of its geothermal potential. In this paper, we report out the results of the geophysical study performed in the Ringelbach site on an exhumed transition zone in the Vosges Mountain, on the western flank of the Rhine graben. We focused on electrical and acoustic properties of the transition zone as they are the main physical parameters usually assessed with the help of geophysical methods during the exploration phase of a geothermal project.

## 2. RINGELBACH ANALOGUE SITE

The Ringelbach site is located on the eastern side of the Vosges Massif (NE France) (Figure 1) and consists of Hercynian porphyritic granite overlaid by a residual cover of Triassic sandstones. The granite is medium-grained, and consists of amphibole, biotite, quartz, plagioclase, and K-feldspar with minor titanite (Wyns, 2012). The Triassic cover gently dips towards the north; it essentially consists of thick, hard medium-grained sandstones, alternating with a series of thin quartz-silty clay layers.

Three boreholes (F-HUR, F-HEI and F-HEI2) were drilled within the area for hydrogeological purpose (Figure 1). A detailed petrographic description of the borehole samples is given in Wyns (2012). The lithological profiles of F-HUR and F-HEI are summarized in Figure 2. Neither of the two 150-m deep boreholes (F-HUR, F-HEI) reach the fresh granite.

Darnet et al.

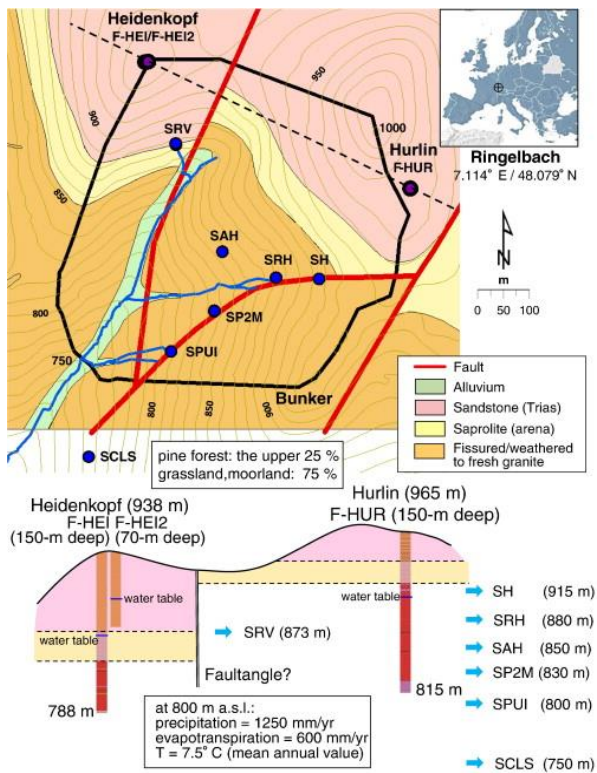

Figure 1: Topographic and geologic map of the Ringelbach catchment with the locations of the springs (blue circles) and boreholes (violet circles). The dotted line indicates the position of the geological section presented below. F-HUR borehole: 150 m deep, through sandstone (0–28 m), granitic arena (28–48 m) and fissured granite. F-HEI borehole: 150 m deep, through sandstone (0–74 m), granitic arena (74–101 m) and fissured granite.

These two deep boreholes indicate that sandstone in the top of the Ringelbach site covers a 20- to 35-m-thick upper layer of clayey saprolite developed in granite. Saprolite is characterized by total alteration of ferroan-magnesian minerals liberating iron to form iron oxi-hydroxides, and by argilization of plagioclase. The induration and foliation of this truncated arena layer increase at its bottom. Underneath is a very thick fissured layer in which iron hydroxide content and argilization of plagioclase progressively decrease. This succession of sub-horizontal layers is interpreted as a truncated pre-Triassic stratified weathering profile of the Hercynian granite (Wyns, 2012), as also proposed for other European Hercynian regions (Wyns et al., 2004).

Paragenetic sequence of fracture fillings and wallrock alteration provide evidence of an early cataclasis stage associated with micro-quartz cementation followed by carbonates and barite. Preliminary microthermometric study of fluid inclusions in carbonates (F-HUR 138 m) allows measuring homogeneisation temperatures of 120-160°C; these temperatures are very similar to those obtained at Soultz (EPS1 borehole) or in granite of the Waldhambach quarry (Dezayes and Lerouge, 2019). These temperatures suggest hydrothermal conditions at maximum burial cover/basement interface reached just before the graben opening.

Two main faults divide the basin into three blocks (Bunker, Hurlin, and Heidenkopf), which are progressively downthrown from southeast to northwest (Figure 1). Within each block, periglacial and postglacial weathering and erosion processes generated a more recent regolith on the slope surfaces, which is typically only a few meters thick.

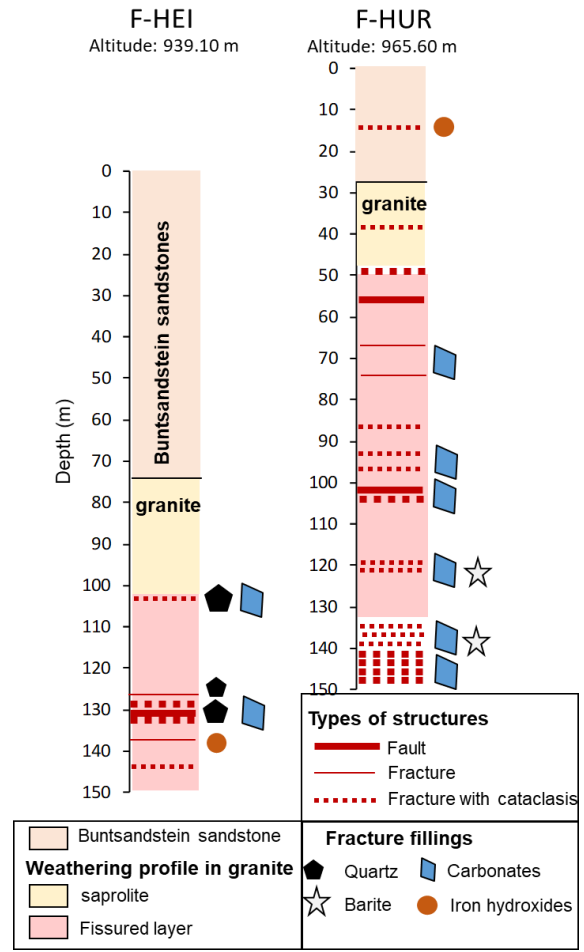

Figure 2: Lithological profiles of F-HEI (left) and F-HUR (right) boreholes, including paleoweathering zones in granite, and major structures and fracture fillings in granite modified from Wyns (2012)

## 3. ELECTRICAL PROPERTIES

### 3.1 Core-scale measurements

Electrical formation factor and surface conductivity measurements have been performed on Ringelbach's granitic core samples taken from the F-HUR and F-HEI wells (Belghoul, 2007). It shows that the higher porosity, the higher surface and pore fluid conductivity are (Figure 3). Since the amount of alteration clays controls the rock surface conductivity (Revil et al, 1998), we can conclude that the higher the degree of alteration of the granite is, the higher the conductivity is. In addition, the higher the porosity is, the higher the permeability of the samples is (Figure 4). Porous and permeable altered granitic formations found within the transition zone are therefore likely to exhibit elevated electrical conductivity compared to unaltered and tight granite.

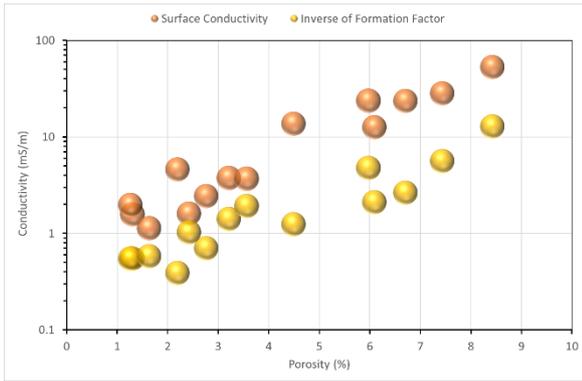

**Figure 3:** Surface (orange) and pore conductivity (yellow) of altered granite core samples extracted from the Ringelbach transition zone (Belghoul, 2007).

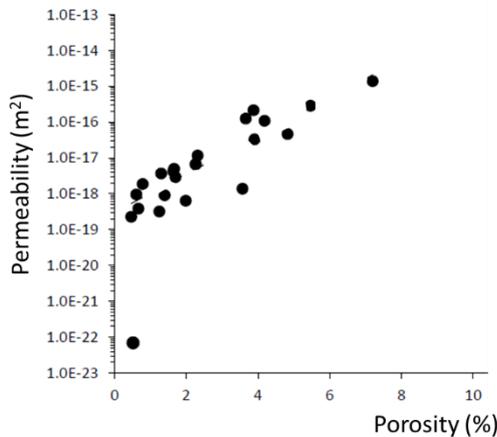

Figure 4: Permeability of Ringelbach's core samples as a function of the porosity of the samples (Belghoul, 2007).

### 3.2 Borehole logging

Figure 5 display the logged resistivity data in the F-HUR well. Resistivity logs show overall high values (> 250 Ohm.m) over the crystalline basement but also a high degree of variability (250 – 6000 Ohm.m) caused by variations in the degree of alteration/fracturing of the granite, as evidenced by the strong correlation between resistivity readings and the rock competency. The close relationship between elevated porosity/permeability and high electrical conductivity observed at the scale of the core samples can therefore be transposed to the scale of the transition zone (~100m thick, here).

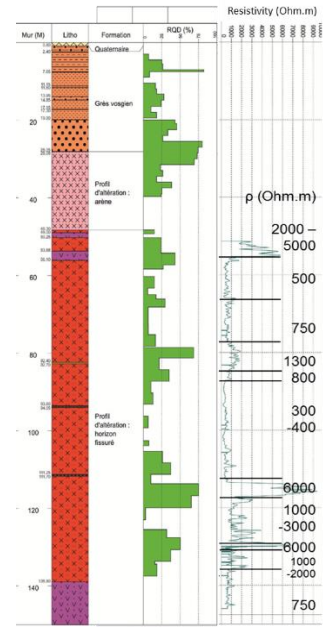

**Figure 5:** Logged resistivity data from F-HUR well. RQD log is a proxy for the competency of the rock. Black lines marks the limits of units with similar electrical properties. Values in black provides estimates of electrical properties per unit.

### 3.3 Field scale CSEM measurements

To characterize the large-scale electrical resistivity distribution of the transition zone, we acquired a 3D Controlled Source Electro-Magnetic (CSEM) survey over a 1000m x 1000m area cutting through the transition zone (from the Triassic sedimentary cover deep down into the crystalline basement). It used 50 recording stations and one vertical transmitting loop (Figure 6). The 3D resistivity cube obtained from the inversion of such data shows that the conductive anomaly associated with altered and fractured zone extends well over 200m depth into the basement and is laterally extensive (Figure 7). It also shows that the depth of this conductive zone below the base of the sedimentary cover varies laterally (from zero to 200m depth), most likely due to a different alteration history of the granite.

Darnet et al.

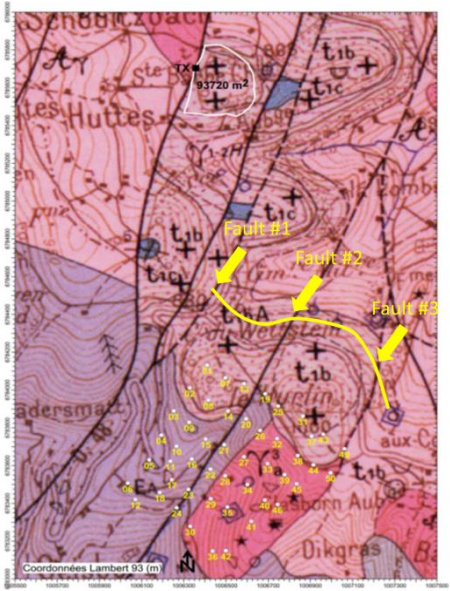

**Figure 6: 3D CSEM receiver grid (yellow dots) and transmitter (white loop) and 2D seismic line (yellow line) deployed over the Ringelbach area. Background map is the geological map of the area (t1b is Permian sandstone, purple and red gamma is granite)**

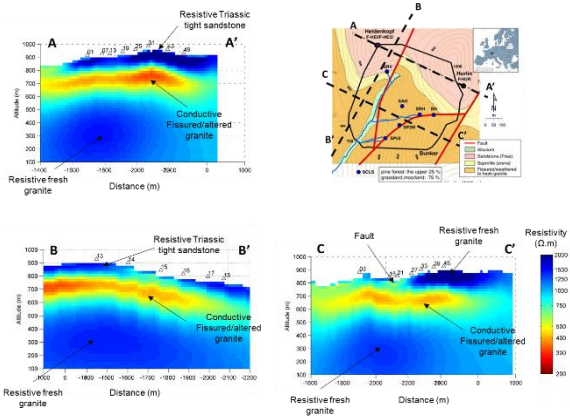

**Figure 7: Left: 3D resistivity cubes (top: strike and bottom: dip lines) obtained from the inversion of 3D Controlled-Source Electro-Magnetic data acquired over the Ringelbach area. Right: Geological map (top) and cross-section (bottom) of the Ringelbach area.**

## 4. ACOUSTIC PROPERTIES

### 4.1 Core-scale measurements

P-wave velocity measurements (Figure 8) have been performed on Ringelbach's core samples taken from the F-HUR and F-HEI wells (Belghoul, 2007). It shows that the higher porosity, the lower the P-wave velocity is. As for the electrical conductivity, we can therefore expect that porous and permeable altered granite found within the transition zone exhibit lower P-wave velocity than unaltered and tight granite.

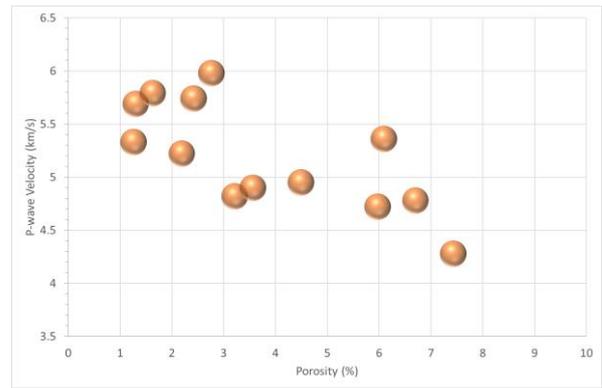

**Figure 8: P-wave velocity of altered granite core samples extracted from the Ringelbach transition zone (Belghoul, 2007).**

### 4.2 Borehole logging

Figure 9 display the logged sonic velocities in the F-HUR well. Sonic logs show overall high values (> 4500 m/s) over the crystalline basement but also a high degree of variability (3000 – 5500 m/s) caused by variations in the degree of alteration/fracturing of the granite, as evidenced by the strong correlation between sonic readings and the rock competency. The close relationship between elevated porosity/permeability and low P-wave velocity observed at the core scale can therefore be scaled up to the size of the transition zone (~100m thick, here).

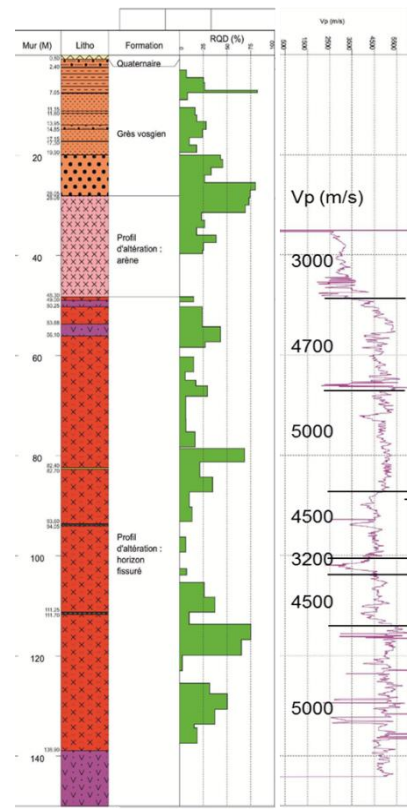

**Figure 9: Logged sonic data from F-HUR well. RQD log is a proxy for the competency of the rock. Black lines marks the limits of units with similar acoustic properties. Values in black provides estimates of P-wave velocity per unit.**

## 3.3 Field scale seismic measurements

To characterize the large-scale P-wave velocity distribution of the transition zone, we acquired a 2D seismic reflection profile over the Triassic sedimentary cover cutting through three fault zones identified in the area (Figure 6). A clear velocity drop from 4000-5000 m/s down to 3500-4000 m/s is observed at the top of the crystalline basement (within the first 100m) when crossing these faults (Figure 10). Similarly, reflectors associated to layering within the basement fade away when crossing these fault zones, possibly indicating the presence of strongly altered and fractured granite. Interestingly, numerous and continuous reflectors are present within the basement, possibly caused by some layering in the alteration of the granite, as evidenced by the layering observed on the sonic logs in the exploratory boreholes.

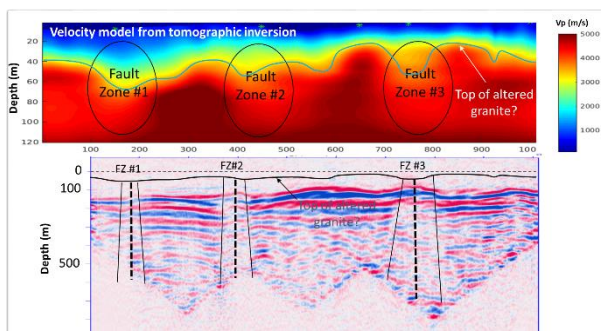

**Figure 10:** Top: Velocity model derived from the travel-time inversion of first breaks. Bottom: stacked migrated section in depth

## 5 DISCUSSION

### 5.1 Surface versus pore electrical conductivity

Waxman and Smits' (1968) equation can used to describe the relationship between surface, pore and total electrical conductivity of a clay-rich porous media as follows:

$$\sigma_0 = \frac{\sigma_w}{F} + \sigma_s$$

with $\sigma_0$ = rock conductivity, $\sigma_s$ = surface conductivity, $\sigma_w$ = fluid conductivity and $F$ = rock formation factor.

We computed the total rock conductivity at 10 and 200 degC as function of the fluid salinity for altered and unaltered Ringelbach granite (Figure 11) using Sen and Goode (1992) relationship for the brine conductivity and the following constants derived from the core sample analysis (section 3.1):

|  | Porosity | Formation Factor | Surface Conduction | Permeability |
|---|---|---|---|---|
| **Fresh granite** | 1.8% | 1900 | 2.3 mS/m | $10^{-18}$ m$^2$ |
| **Altered granite** | 5.3% | 480 | 20 mS/m | $10^{-16}$ m$^2$ |

The less saline/colder the fluid, the more important the surface conduction is. For the Ringelbach case, groundwater is fresh (salinity ~100 mg/L) and as a consequence, the total conductivity of the granite is mainly due to surface conduction and hence a proxy of mineral alteration but not indicative of rock porosity. Conversely, the more saline/hotter the fluid, the more important pore fluid conduction is. For deep and hot geothermal brines like found in Soultz-sous-Forêts in the Rhine graben (salinity of 100 g/L and temperature of 200 degC), pore fluid conductivity is dominating and the total rock conductivity is indicative of the rock porosity and hence potentially a proxy of the rock permeability.

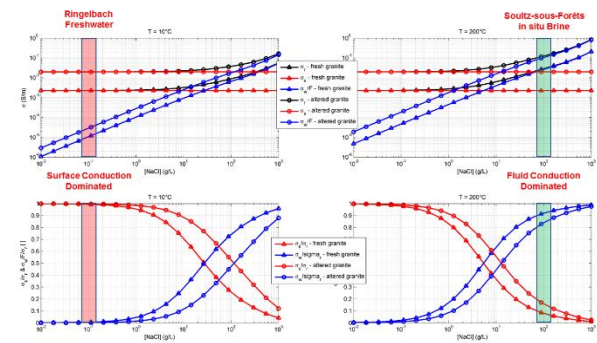

**Figure 11:** Top left: surface (red), pore fluid (blue) and total (black) conductivity of unaltered (triangle) and altered (circle) granite at 10 degC as a function of the brine salinity. Bottom left: ratio between the surface (red) and pore fluid (blue) versus the total rock conductivity for of fresh (triangle) and altered (circle) granite at 10 degC as a function of the brine salinity. Top right: surface (red), pore fluid (blue) and total (black) conductivity of unaltered (triangle) and altered (circle) granite at 200 degC as a function of the brine salinity. Bottom right: ratio between the surface (red) and pore fluid (blue) versus the total rock conductivity for of fresh (triangle) and altered (circle) granite at 200 degC as a function of the brine salinity.

### 5.2 Electrical and acoustic properties of deep geothermal reservoirs

To validate the aforementioned electrical conductivity model for deep geothermal reservoirs, we present here electricial resistivity measurements made in the Soulzs-sous-Forêts granitic basement (Figure 12). As predicted, altered zones are electrically more conductive than unaltered granite, with a factor ranging from ten to thousand times. Similarly, in the Rittershoffen geothermal project, altered zones at the top of the granitic basement proved to be electrically conductive but notably, the main permeable fault zones coincide with the most electrically conductive zones (Glass et al., 2018). The electrical conductivity of the granitic basement is therefore a parameter of choice to explore for deep porous and potentially permeable fault zones. The challenge remains to deploy electromagnetic techniques capable of remotely detecting and imaging such anomalies, deeply buried underneath a thick sedimentary cover.

Darnet et al.

On acoustic measurements, the most altered zones within the basement in Soultz-sous-Forêts show significant decrease in P-wave and S-wave velocities (up to 10%). However, the less altered zones do not exhibit any significant anomalies, most likely due to the limited sensitivity of elastic properties to alteration. To compensate for that, a joint analysis of electrical and elastic properties is necessary. Indeed, when crossploting logged P-wave velocity as a function of resistivity, we can observe that the less altered granites have similar P-wave velocities but that their resistivity varies greatly (by a factor 100) allowing to discriminate between the different alteration facies. Here also, the challenge remains to deploy seismic techniques capable of remotely detecting and imaging such anomalies.

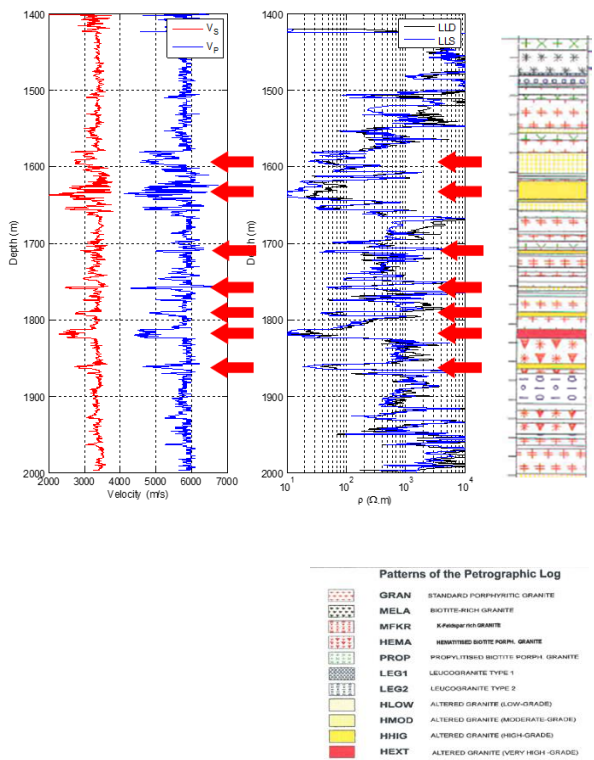

**Figure 12: P-wave (blue) and S-wave (red) logs (left), shallow (blue) and deep (black) laterolog (middle) and lithological log (right) in the Soultz-sous-Forêts GPK1 well. Most altered granitic facies are marked with red arrows.**

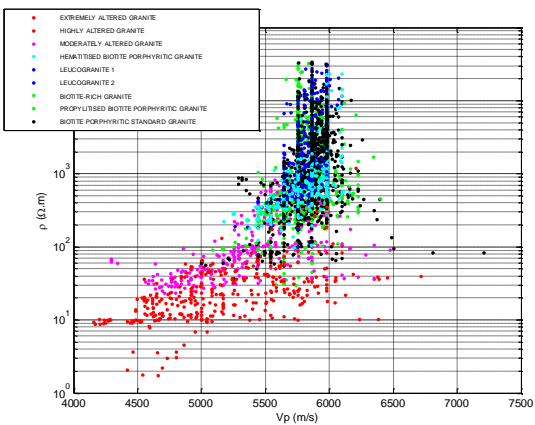

**Figure 13: Logged resistivity as a function of the P-wave velocity of the granitic basement found in the Soultz-sous-Forêts GPK1 well. Colors indicate the different alteration facies.**

## CONCLUSION

Our study of an exhumed transition zone in the Vosges Mountain shows that altered porous and potentially permeable granite targeted in deep geothermal exploration has a clear signature on electrical conductivity and acoustic measurements. It also showed that the best discrimination comes from the joint interpretation of these datasets. The challenge remains to develop and deploy electromagnetic and seismic imaging techniques having sufficient resolution to detect and image the targets of interest, usually deeply buried underneath a thick sedimentary cover.

## ACKNOWLEDGEMENT

We would like to thank Albert Genter and Vincent Maurer from Electricité de Strasbourg – Géothermie for fruitful discussions on logging data in deep geothermal projects.